\documentclass[twocolumn,english,prl]{revtex4}
\usepackage[latin9]{inputenc}
\usepackage{amsmath}
\usepackage{amssymb}

\makeatletter
\@ifundefined{textcolor}{}
{%
 \definecolor{BLACK}{gray}{0}
 \definecolor{WHITE}{gray}{1}
 \definecolor{RED}{rgb}{1,0,0}
 \definecolor{GREEN}{rgb}{0,1,0}
 \definecolor{BLUE}{rgb}{0,0,1}
 \definecolor{CYAN}{cmyk}{1,0,0,0}
 \definecolor{MAGENTA}{cmyk}{0,1,0,0}
 \definecolor{YELLOW}{cmyk}{0,0,1,0}
 }

\usepackage{dcolumn}\usepackage{bm}\usepackage{babel}

\@ifundefined{textcolor}{}{
 \definecolor{BLACK}{gray}{0}
 \definecolor{WHITE}{gray}{1}
 \definecolor{RED}{rgb}{1,0,0}
 \definecolor{GREEN}{rgb}{0,1,0}
 \definecolor{BLUE}{rgb}{0,0,1}
 \definecolor{CYAN}{cmyk}{1,0,0,0}
 \definecolor{MAGENTA}{cmyk}{0,1,0,0}
 \definecolor{YELLOW}{cmyk}{0,0,1,0}
 }

\makeatother

\usepackage{babel}
\begin{document}

\title{Comment on \char`\"{}Foundation of Statistical Mechanics under Experimentally
Realistic Conditions\char`\"{}}

\maketitle
In a recent influential letter \cite{Reimann08}, Reimann demonstrates
equilibration of isolated quantum many-body systems from non-equilibrium
initial states by proving the following result: 
\begin{equation}
\sigma_{A}^{2}\equiv\overline{[Tr\{\rho(t)A\}-Tr\{\rho_{eq}A\}]^{2}}\leq\Delta_{A}^{2}\sum_{n}\rho_{nn}^{2}(0),\label{eq:1}
\end{equation}
where $\rho_{mn}(0)=\langle m|\rho(0)|n\rangle$ denotes the initial
density matrix of the system under the eigenbasis $\left\{ |n\rangle\right\} $
of the Hamiltonian $H$ with eigenenergy $\left\{ E_{n}\right\} $,
$\rho(t)=\sum_{mn}\rho_{mn}(0)e^{i\left(E_{n}-E_{m}\right)t}\left\vert m\right\rangle \left\langle n\right\vert $
(setting $\hbar=1$) is the density matrix at time $t$, $\overline{\cdots}$
denotes the time average, $\rho_{eq}=\sum\rho_{nn}(0)|n\rangle\langle n|$
, and $\Delta_{A}$ is the working range of the observable $A$ \cite{Reimann08}.
By assuming the conditions that 1) the system's initial state has
a broadly spread energy level population such that the quantity $\sum_{n}\rho_{nn}^{2}(0)$
is exponentially small with the number of degrees of freedom, and
2) the observable $A$ has a bounded $\Delta_{A}$, the author shows
that Eq. (1) implies that the time averaged variance of the observable
$A$ from its equilibrium value is exponentially small. This is a
strong result. Unfortunately, we show here that the author's proof
of the major result given by Eq. (1) is flawed and invalid for generic
many-body systems under experimentally realistic conditions.

There is a critical loophole in the proof of Eq. \ref{eq:1} in Ref.
\cite{Reimann08}. In simplifying the summation $\sigma_{A}^{2}=\sum_{j,k,m,n}^{\prime}\tilde{A}_{jk}\rho_{kj}\tilde{A}_{mn}\rho_{mn}(0)\overline{e^{i[E_{j}-E_{k}+E_{m}-E_{n}]t}}$
in Eq.(11) of \cite{Reimann08} ($\sum^{\prime}$ means $j\neq k$
and $m\neq n$ in the summation), the author drops all the terms except
for the ones with $j=n$ and $k=m$ by exploiting a questionable {}``fact\textquotedblright{}
that $\overline{e^{i[E_{j}-E_{k}+E_{m}-E_{n}]t}}$ vanishes for all
the other terms under generic Hamiltonians. This {}``fact\textquotedblright{}
is not valid for many-body systems under experimental realistic conditions.
To see this clearly, let us first look at systems with quadratic Hamiltonians,
which represent an important class of physical systems where the study
of equilibration has recently attracted great theoretical and experimental
interest (see Refs. {[}2-4{]} in Reimann's paper). The quadratic systems
can always be diagonalized as a set of non-interacting quasi-particle
modes. The system has eigenenergy $E_{n}=\sum_{i}n_{i}\omega_{i}$,
where $n_{i}$ is the quasi-particle number of the $i^{th}$ mode
with frequency $\omega_{i}$. The condition $E_{j}-E_{k}+E_{m}-E_{n}=\sum_{i}(j_{i}-k_{i}+m_{i}-n_{i})\omega_{i}=0$
does not necessarily require $j_{i}=n_{i}$ and $k_{i}=m_{i}$. For
instance, $E_{j}-E_{k}+E_{m}-E_{n}=0$ can be easily satisfied with
$j_{i}+m_{i}=k_{i}+n_{i}$ ($j_{i}\text{, }m_{i}\text{, }k_{i}\text{, }n_{i}$
are non-negative integers), which already leads to an exponentially
large number of other solutions with non-vanishing $\overline{e^{i[E_{j}-E_{k}+E_{m}-E_{n}]t}}$.
Keeping only the terms with $j=n$ and $k=m$ in the summation drops
out many (exponentially large in the number of modes) other terms,
which leads to the artifact that $\sigma_{A}^{2}$ is bounded by an
exponentially small quantity.

One may argue that the special spectrum of the quadratic Hamiltonians
leads to the above problem and hope that the terms with $\left(j,k\right)\neq\left(n,m\right)$
can still be dropped for generic non-integrable Hamiltonians. This
is not true and we show here that dropping the terms with $\left(j,k\right)\neq\left(n,m\right)$
is unjustified for any many-body Hamiltonians under experimentally
realistic conditions. Note that for a large system with $N$ particles,
typically its energy spread $\Delta E$ scales up linearly with the
particle number $N$, but the number of energy levels scales up exponentially
with N. So, for many-body system, its energy spectrum is always exponentially
dense, with the energy difference between adjacent levels going down
exponentially to zero. For any systems of physical relevance, the
time window $T$ for doing the average over $\overline{e^{i[E_{j}-E_{k}+E_{m}-E_{n}]t}}$
is finite. For instance, $T$ cannot exceed the age of the universe.
However, because of the exponentially dense energy spectrum of many-body
systems, for any given small energy interval $\epsilon\sim1/T$, there
are always exponentially large number of near resonant terms with
$\left\vert E_{j}-E_{k}+E_{m}-E_{n}\right\vert <\epsilon$ which give
non-vanishing $\overline{e^{i[E_{j}-E_{k}+E_{m}-E_{n}]t}}$, and all
these terms cannot be dropped in the summation. In other words, to
resolve the energy difference between adjacent levels for many-body
systems, even for just hundreds of particles, the time window $T$
required for the average $\overline{\cdots}$ needs to be far beyond
the age of the universe, which is clearly of no physical relevance.
So the claim in Ref. \cite{Reimann08} for derivation of equilibrium
under experimentally realistic conditions is unjustified.

Although there are many recent works showing either numerically or
experimentally that some isolated quantum systems appear to relax
to $\rho_{nn}(0)$, much more complicated requirements on initial
state and/or Hamiltonian are usually required for relaxation. The
two conditions in Ref. \cite{Reimann08} are too simple and insufficient
to guarantee relaxation due to the aforementioned loophole in the
proof.

This work was supported by the NBRPC (973 Program) 2011CBA00300 (2011CBA00302),
the DARPA OLE program, the IARPA MUSIQC program, the ARO and the AFOSR
MURI program.

Zhe-Xuan Gong and L. -M. Duan \\
Department of Physics and MCTP, University of Michigan, Ann Arbor,
Michigan 48109 and \\
Center for Quantum Information, IIIS, Tsinghua University, Beijing,
China


\begin{thebibliography}{References}
\bibitem{Reimann08} P. Reimann, Phys. Rev. Lett. 101, 190403 (2008). \end{thebibliography}
\end{document}